\documentclass[pre,aps,superscriptaddress,twocolumn]{revtex4-1}
\usepackage{graphicx} 
\usepackage{xcolor,hyperref}
\usepackage{amsmath,amsfonts,amssymb}
\usepackage{subcaption}
\usepackage{graphicx}
\usepackage{float}
\usepackage[normalem]{ulem}

\usepackage{soul}

\begin{document}

\title{Coarse-graining amorphous plasticity: impact of rejuvenation and disorder}

\author{Botond Tyukodi}
\affiliation{Department of Physics, Babe\c{s}-Bolyai University, 400084 Cluj-Napoca, Romania}
\affiliation{Martin Fisher School of Physics, Brandeis University, Waltham, Massachusetts 02454, USA}
\author{Armand Barbot}
\affiliation{ PMMH, CNRS, ESPCI Paris, Sorbonne Universit\'e, Universit\'e de Paris, Universit\'e PSL, F-75005 Paris France}
\author{Reinaldo Garci\'a-Garci\'a}
\affiliation{Departamento de Física y Matemática Aplicada, Facultad de Ciencias, Universidad de Navarra, Pamplona 31008, Spain}
\author{Matthias Lerbinger}
\affiliation{ PMMH, CNRS, ESPCI Paris, Sorbonne Universit\'e, Universit\'e de Paris, Universit\'e PSL, F-75005 Paris France}
\author{Sylvain Patinet}
\affiliation{ PMMH, CNRS, ESPCI Paris, Sorbonne Universit\'e, Universit\'e de Paris, Universit\'e PSL, F-75005 Paris France}
\author{Damien Vandembroucq}
\affiliation{ PMMH, CNRS, ESPCI Paris, Sorbonne Universit\'e, Universit\'e de Paris, Universit\'e PSL, F-75005 Paris France}


\begin{abstract}
The coarse-graining of amorphous plasticity from the atomistic to the mesoscopic scale is studied in the framework of a simple scalar elasto-plastic model. Building on recent results obtained on the atomistic scale, we discuss the interest {in a disordered landscape-informed} threshold disorder to reproduce the physics of amorphous plasticity. We show that accounting for a rejuvenation scenario allows us to reproduce quasi-quantitatively the evolution of the mean local yield stress and the localization behavior. We emphasize the crucial role of two dimensionless parameters: the relative strength of the yield stress disorder with respect to the typical stress drops associated {with a plastic rearrangement}, and the age parameter characterizing the relative stability of the initial glass with respect to the rejuvenated glass that emerges upon shear deformation.
\end{abstract}

\maketitle

\section{Introduction}

The plastic deformation of glasses and amorphous solids is a complex phenomenon. Dependence of the mechanical behavior (e.g. brittle or ductile) on preparation, avalanches and shear-banding are among its classical hallmarks. These features come together with a cohort of other puzzling properties: aging or rejuvenation~\cite{Debenedetti-PRL00,Egami-NatComm17,Egami-ActaMater18,Patinet-PRE20-rejuvenation}; plasticity-induced anisotropy and Bauschinger effect~\cite{RVTBR-PRL09,Procaccia-PRL10,Patinet-PRL2020-Bauschinger}; memory effects under cyclic loading~\cite{Sastry-PRL14,Keim-RMP19}.


The understanding of the mechanical properties of glasses and amorphous materials has thus motivated an increasing amount of studies in the last decades~\cite{Schuh-ActaMat07,RTV-MSMSE11,Bonn-RMP17,Tanguy-CRP21}. In the absence of crystalline structure and thus of dislocations, it has appeared in particular that plastic deformation in amorphous materials results from {a series} of localized rearrangements of the amorphous structure, Shear Transformations~\cite{Argon-ActaMet79,FalkLanger-PRE98} that interact through the surrounding elastic medium~\cite{BulatovArgon94a,BulatovArgon94b, BulatovArgon94c}. The associated complex mechanical behavior naturally involves a wide range of time and length scales~\cite{RTV-MSMSE11}. 

In parallel to the increasing number of atomistic simulations developed to study and characterize these various features, the last decades have seen the development of over-simplified lattice elasto-plastic models operating at a mesoscopic scale (see the recent review of Nicolas et al.\cite{Nicolas-RMP18} and references therein). Relying on the coupling of a local threshold dynamics and a {long-range} elastic interaction induced by the local plastic rearrangements of the amorphous structure, these models were able to reproduce most of the complexity of amorphous plasticity despite the drastic reduction of degrees of freedom. In the same spirit as the Ising-like models for magnetism or the shell models for fluid turbulence, one may consider that {mesoscopic Elasto-Plastic Models} (EPMs) belong to the class of models recently  coined by Bouchaud~\cite{Bouchaud-EPN19} as \emph{metaphoric}.  Such models  do not attempt {to describe reality precisely} or necessarily rely on very plausible assumptions. {Instead}, they aim to illustrate  non-trivial {mechanisms}, the scope of which goes much beyond the specifics of the model itself.

Conversely, mesoscopic {EPMs} can also be {considered} as an important step in the framework of a multiscale modelling of amorphous plasticity, operating at an intermediate scale between atomistic simulations and constitutive models at continuous scale~\cite{van-der-Giessen-MSME20}. Such a multiscale strategy requires {identifying} the parameters to be transferred to the upper scale and the calibration between {the microscopic and mesoscopic scales}. Driven by the rapid development of new characterization techniques of local shear  transformations~\cite{Puosi-SM15,Albaret-PRE16,PVF-PRL16,Patinet-PRE18,Richard-PRM20}, this difficult question of {a} quantitative connection between atomistic and mesoscopic scale has very recently been addressed in Refs.{~\cite{Patinet-CRP21,Pardoen-IJP21,Liu-Martens-PRL21,Patinet-Acta22}.}

{At first sight, a complete quantitative agreement between atomistic and mesoscopic scales may appear elusive since it requires an ever-increasing number of parameters. Although not perfect, the above-cited } works impressively reproduce several key aspects of amorphous plasticity upon quasi-static or finite shear rate loading: stress-strain curves, flow curves, {creep~\cite{Liu-Martens-PRL21}} but also some of the microscopic and macroscopic fluctuations, as well as non-trivial emergent behaviors such as the Bauschinger effect{~\cite{Patinet-CRP21}}. 

Here we intend to discuss coarse-graining in a slightly different spirit. Sticking to the most simple scalar version of a mesoscopic elastoplastic model, we highlight the importance of accounting for the structure and evolution of the underlying disordered glassy landscape. In particular, we explore the rejuvenation scenario proposed in Ref.~\cite{Patinet-PRE20-rejuvenation} which states that upon loading, the glass is gradually transformed through local rearrangements into a new glass phase with plastic properties that are close to that of an inherent state of a supercooled liquid, close to the mode coupling transition.

Building on recent atomistic studies~\cite{Patinet-PRE18,Patinet-PRE20-rejuvenation} we try to understand to what extent we can reproduce the most salient features of AQS (Athermal Quasistatic) simulations in the framework of lattice models with the sole knowledge {of the initial statistical properties grained from atomic configurations.} In particular, from the success, limitations and even failures of the different hypotheses that can be developed within this approach, we try to get more insight into the structure and evolution of the model's Potential Energy Landscape (PEL) glasses upon shearing.

In the following, we first summarize in section~\ref{section:AQS} recent results of AQS simulations of a simple binary model glass~\cite{PVF-PRL16,Patinet-PRE18,Patinet-PRE20-rejuvenation}. In particular, we show the contrasting localization behavior depending on glass preparation {and discuss} the statistics of local plastic thresholds in the initial state. In section~\ref{section:Mesomodel} we give a short presentation of the mesoscopic {EPM}. In section~\ref{section:upscaling}, we present our up-scaling strategy. In section~\ref{section:ESL} and \ref{section:GQ} we present and discuss comparisons of atomistic and mesoscopic results obtained for a poorly annealed glass and a gradually quenched glass characterized by a shear-banding behavior. We finally discuss in section~\ref{section:Discussion} different perspectives about the coarse-graining of amorphous plasticity.

\section{Atomistic simulations of model glasses}\label{section:AQS}
\subsection{Preparation}

Three different two-dimensional binary Lennard-Jones (LJ) model glasses obtained with contrasting thermal histories are considered in the present article. For each glass{,} 100 samples of 10000 atoms are prepared. The atomistic models and the protocols of preparation used here have been presented in {detail} in Refs.~\cite{Patinet-PRE18,Patinet-PRE20-rejuvenation}. Periodic boundary conditions are applied to the simulation box. In the following{,} all the physical quantities are given in LJ units. The binary system is prepared in the liquid state (NVT ensemble) at high temperature, then cooled down and equilibrated to a supercooled liquid state at $T=T_p$. Then the
supercooled liquid at parent  temperature $T_p$ is either abruptly quenched by energy minimization or gradually quenched at a slow rate down to a {low-temperature} glassy state ($T\approx0.03$) before it is also abruptly quenched by energy minimization.

A first glass is obtained from the instant quench of a supercooled liquid equilibrated at $T_p=0.38$. This temperature is close to the estimated temperature of the Mode Coupling Transition of our system $T_{MCT} = 0.373\pm0.01$~\cite{Patinet-PRE20-rejuvenation}. This glass {, which we will refer to as MCT in the following,} will be used as a proxy of the glass structure at Mode Coupling Transition.

A second glass is obtained from the instant quench of a supercooled liquid equilibrated at $T_p=0.351$. This temperature is still significantly higher than {the numerical temperature of glass transition, estimated at $T_G\approx0.31$}. This glass which we will refer to in the following as ESL (for Equilibrated Supercooled Liquid) will be used as a typical poorly annealed glass.

A third glass is obtained from the gradual quench of a supercooled liquid equilibrated at $T_p=0.351$ and {slowly quench over a period of $10^6$}. This glass which we will refer to in the following as GQ (for Gradual Quench) will be used as a typical well-annealed glass. This statement is obviously to be considered in the context of conventional molecular dynamics, i.e. in the absence of accelerated equilibrium procedures such as {swap Monte Carlo algorithm}~\cite{Berthier-JStat19}.

\subsection{Athermal Quasistatic Shear (AQS) loading}
The glass samples are deformed in simple shear geometry in Athermal Quasi-Static conditions, i.e. by series of elementary deformation operations consisting of one small incremental affine global shear step followed by an energy
minimization step~\cite{Maloney-PRL04a,Maloney-PRE06}. The shearing protocol is applied up to a global shear strain $\gamma=5$. {Local strain fields are computed by integration using Hencky's finite strain definition.} Details about the computation of the local stress and strain fields can be found in Ref.~\cite{Patinet-PRE18}.


\begin{figure}
\begin{subfigure}[b]{0.95\columnwidth}
\includegraphics[width=\columnwidth]{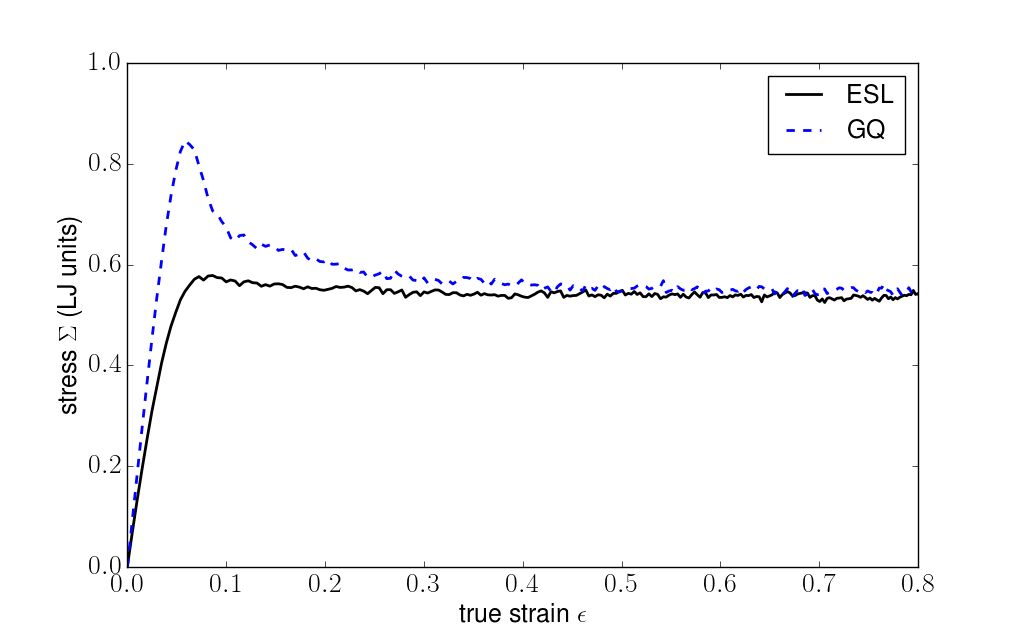}\caption{}
\end{subfigure}
\begin{subfigure}[b]{0.485\columnwidth}
\includegraphics[width=\columnwidth]{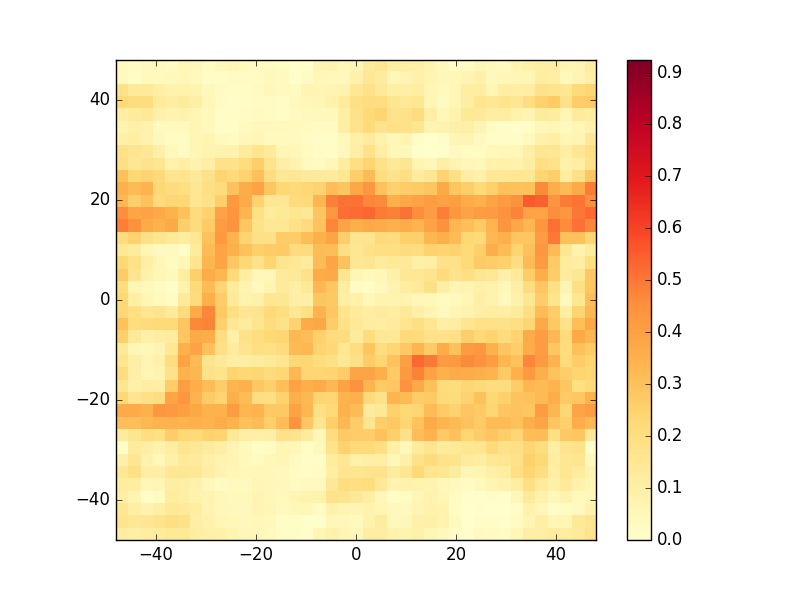}\caption{}
\end{subfigure}
\begin{subfigure}[b]{0.485\columnwidth}
\includegraphics[width=\columnwidth]{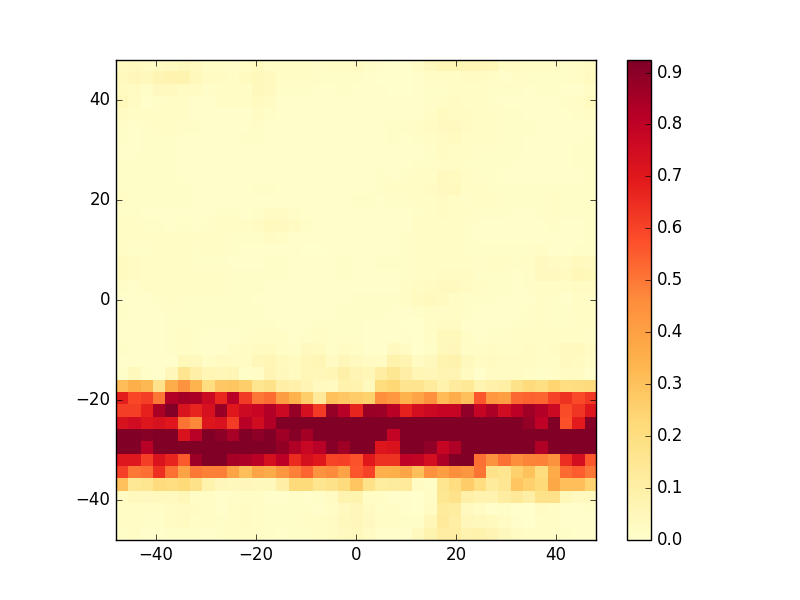}\caption{}
\end{subfigure}
  \caption{Plastic behavior obtained at atomistic scale for fast and slow quenched glasses ESL and GQ respectively \--- a) Stress-strain curves (after averaging over 20 realizations) \--- Plastic strain field after 20\% of total shear deformation in b) the fast quenched glass ESL and c) the gradually quenched glass GQ. A shear-band is clearly visible in the latter case. }
  \label{AQS-results}
\end{figure}

In Fig.~\ref{AQS-results}a we show the stress curves obtained for the {soft glass ESL }and the hard gradually quenched glass GQ. While ESL shows a quasi-monotonous stress strain curve, GQ exhibits a stress peak followed by a softening branch. Eventually, independently on the initial state, the two glasses reach the same stress plateau. In Fig.~\ref{AQS-results}b and c, we show strain maps of obtained after deforming  ESL and GQ by $20\%$. While mild spatial fluctuations are observed for ESL, a well defined shear band is clearly visible for GQ.

\subsection{Local Yield Stress}
 The propensity of the model glass for plastic deformation is estimated through the frozen matrix method exposed and detailed in Refs.~\cite{PVF-PRL16,Patinet-PRE18}. A circular region of radius $\xi$ is isolated in the model glass, with a surrounding shell of width $2R_c$, equal to twice the cut-off length of the interaction potential. The size $\xi=5\sigma$ chosen for the radius of the inner core is such that continuum elasticity is obeyed but with significant fluctuations of the local moduli and anisotropy~\cite{Tsamados-PRE09}.  Series of affine steps of shear deformation are imposed to the (frozen) shell.
After each step, only the inner core is allowed to relax. The deformation is applied up to a plastic event that takes place in the core. The associated yield stress $\sigma_c$ and stress drop $\Delta \sigma$ are computed. This operation is repeated in 18 different shearing directions regularly spaced every $10^{\circ}$. Due to the heterogeneity of local elastic properties, the minimum stress to apply along a given direction to get a plastic rearrangement may be obtained for an affine deformation of the frozen outside shell in nay other direction. The final local yield stress $\sigma_c$ in the $xy$ direction is thus defined as the (positive) minimum of the projected stress thresholds obtained in the 18 different directions~\cite{PVF-PRL16}.

We summarize below recent results obtained in Refs.~\cite{PVF-PRL16,Patinet-PRE18} that allow us to characterize the plastic behavior of the three glasses (ESL, GQ, MCT) of interest at the local scale. These results will help us to develop later our up-scaling strategy.


\begin{figure*}
\begin{subfigure}[b]{0.45\textwidth}
\includegraphics[width=\textwidth]{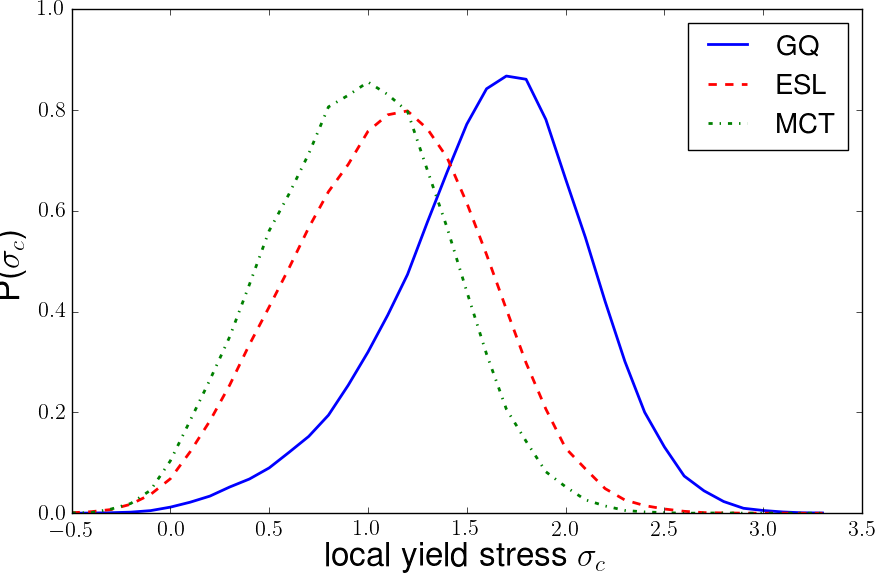}\caption{}
\end{subfigure}
\begin{subfigure}[b]{0.45\textwidth}
\includegraphics[width=\textwidth]{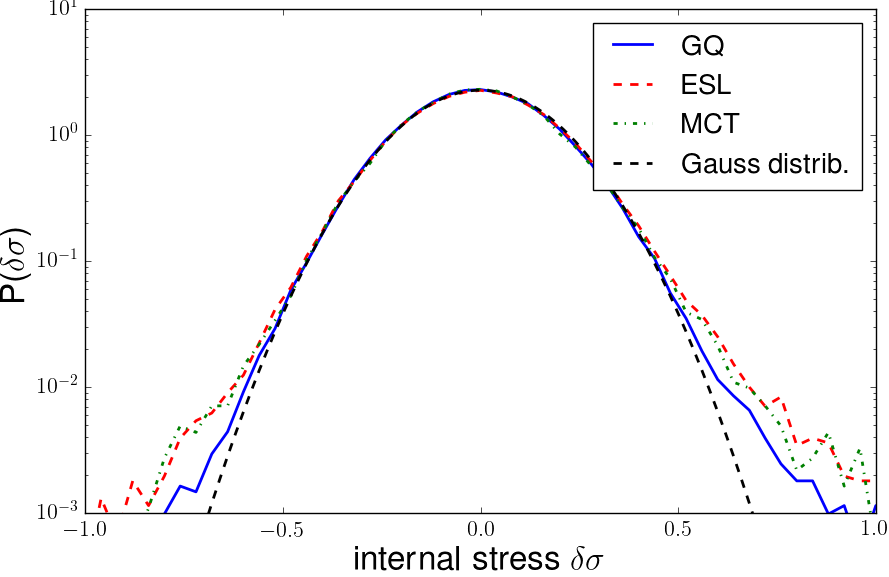}\caption{}
\end{subfigure}

  \caption{Local Yield stress results: a) distribution of local yield stress $\sigma_c$; b) distribution of frozen stress fluctuations $\delta \sigma$}
  \label{LYS-results}
\end{figure*}
  
In Fig.~\ref{LYS-results}a we show the distributions of local yield stresses $P(\sigma_c)$ for the three glasses ESL, GQ and MCT.  As already documented in Refs.~\cite{PVF-PRL16,Patinet-PRE18,Patinet-PRE20-rejuvenation} we see clearly that the younger the glass, the lower the local yield stresses. In other terms, the frozen matrix method developed in Ref.~\cite{PVF-PRL16} allows us to give a microscopic interpretation of the relationship  between age and mechanical stability of glasses. Note that a discrete negative tail is present in all the three cases. This apparently surprising feature results from the existence of a frozen stress field after the quench protocol. If we call $\delta \sigma(\mathbf{r})$ the internal shear stress in a patch located at $\mathbf{r}$ and $\sigma_c(\mathbf{r})$ its local yield stress, mechanical stability simply requires the satisfaction of the inequality $\delta \sigma(\mathbf{r})  < \sigma_c(\mathbf{r})$. We can thus define the local plastic strength $x(\mathbf{r})=\sigma_c(\mathbf{r}) - \delta \sigma(\mathbf{r})$ which, contrary to the local yield stress has to remain positive to ensure mechanical stability. This local plastic strength is nothing but the amount of external shear stress to be applied so that the patch reaches its limit of stability and experiences a plastic rearrangement.

In Fig.~\ref{LYS-results}b we show the distributions of the local internal stress fields $\delta \sigma$ for the three glasses. Interestingly the three distributions fall onto the same Gaussian distribution of width $w_g=0.175$. Despite the presence of a power-law tail for large fluctuations, the quench protocol thus does not seem to alter dramatically the frozen stress fields. We note that these internal stress fluctuations in the initial configuration amount to about $10-20\%$ of the local yield stresses.


  
In Ref.~\cite{Patinet-PRE18} it was moreover shown that the distribution $P(\Delta\sigma)$ of local stress drops $\Delta \sigma$ follows an exponential trend  $P(\Delta \sigma) =  \tau_{max}^{-1} \exp(-\Delta \sigma /\tau_{max})$ with  the older the glass, the larger the local yield stress and the larger the  value of $\tau_{max}$. In addition a parabolic dependence between the local stress drop and the local yield stress was reported: $\overline{\Delta \sigma}= a_0 + a_1 \sigma_c^2$.
  
The frozen matrix method developed in Refs.\cite{Puosi-SM15,PVF-PRL16,Patinet-PRE18} thus appears to be a powerful tool to probe the disordered landscape of glassy material and its evolution upon loading. In the following, we use the new information collected through this method to inform the mesoscopic elastoplastic models.

\section{A simple mesoscopic model of amorphous plasticity}\label{section:Mesomodel}

Simulations at mesoscopic scale are performed using the scalar lattice model developed in Refs.~\cite{VR-PRB11,TPVR-Meso12,Tyukodi-PRL18}. We summarize below its main features. We first recall the definition of the model, we then highlight the crucial role of two dimensionless parameters respectively associated with the amplitude of the elastic interaction and the evolution of the disorder upon loading. We finally introduce the upscaling strategy developed within the present study.


\subsection{Basics of the scalar mesoscopic elastoplastic model}

{\bf Discretization} \--- A spatial discretization is performed at scale $a$ onto a square lattice of size $N\times N$.  Bi-periodic boundary conditions are implemented. The material is considered as elastically homogeneous with a shear modulus $\mu$ and a compressive modulus $K$. In the following we restrict ourselves to a scalar model and only consider the shear component of the stress tensor along the macroscopic shear loading direction. The mechanical state of a given configuration is then given by the shear stress field $\sigma_{ij}$ and the plastic strain field $\varepsilon^p_{ij}$ where $(i,j)$ are the coordinates of the cell. The total strain field is thus the sum of the elastic and plastic strain fields:  $\varepsilon_{ij} =\varepsilon^{el}_{ij} + \varepsilon^p_{ij}$ where by definition $\varepsilon^{el}_{ij}=\sigma_{ij}/2\mu$.

{\bf Local threshold dynamics} \--- A local scalar criterion of plasticity is implemented: stability holds if $\sigma_{ij} < \sigma_{ij}^c$. Whenever the local stress exceeds the local thresholds, the cell experiences a small slip so that the local plastic strain is incremented by a small value $\delta\varepsilon_p$. The latter estimates the plastic strain experienced by a cell associated with a local rearrangement of the amorphous structure. 
The stress landscape of an individual cell thus consists of series of elastic branches of slope $2\mu$ terminating at a (fluctuating) threshold value along the stress scale and separated by (fluctuating) plastic strain increments along the strain scale. Note that here we focus on the mechanical behavior upon a monotonous shear loading and for that reason, we consider only stress thresholds in the forward direction and disregard thresholds in the backward direction.  

{\bf Elastic interaction} \--- Local plastic events take place in cells which are surrounded by other elastic cells. A local plastic strain $\delta e_{pl}$ in an individual cell gives rise to a global plastic strain $\delta\varepsilon_{pl}=\delta e_{pl}/N^2$. At constant total strain this induces a decrease of the elastic strain of the same amount and thus a macroscopic stress drop $\delta \Sigma = 2\mu \delta e_{pl}/N^2$. Moreover it induces a local stress drop within the inclusion and a long range internal stress field of quadrupolar symmetry in the far field. In the case of two-dimensional elasticity, we get for an Eshelby inclusion~\cite{Eshelby1957} of area ${\cal A}=\pi a^2$ experiencing an eigenstrain $\varepsilon_0$:
\begin{align}
   \Delta \sigma^{\mathrm in}(r,\theta) &= -\frac{\mu\varepsilon_0}{1+\mu/K} \;,\\
   \Delta \sigma^{\mathrm out}(r,\theta) &= \frac{\mu\varepsilon_0}{1+\mu/K} \frac{{\cal A} \cos(4\theta)}{\pi r^2} \;.
\end{align}
where the superscripts {\it in} and {\it out} correspond to the inner and outer parts of the inclusion. Note that the amplitude of the local stress drop is directly proportional to the eigenstrain $\varepsilon_0$ and that of the far field is proportional to  the product ${\cal A} \varepsilon_0$ of the inclusion area and the eigenstrain. To account for bi-periodic boundary conditions, the implementation of the Eshelby Green function $G$ is performed in the reciprocal space. Details and discussions about different techniques of implementing this long ranged elastic kernel can be found in Ref.~\cite{TPVR-Meso12,TPRV-PRE16,Tyukodi-PRL18}.

{\bf Disorder} \--- In the initial state, the plastic thresholds $\sigma_{ij}^c$ are drawn from an {\it initial} random distribution $p_i(\sigma_c)$. Whenever a plastic event occurs in a cell, the plastic slip is accompanied by a renewal of the threshold. The new threshold is drawn from a {\it renewal} random distribution $p_r(\sigma_c)$. In the classical {\em disordered} scenario, defined by $p_r=p_i$ the very same distribution is used for the plastic thresholds of the initial state and the plastic thresholds obtained after rearrangements: the materials is left statistically unchanged after a rearrangement . In the {\em rejuvenation} scenario the two distributions $p_r$ and $p_i$ are different. The rationale behind this hypothesis early discussed in Ref.~\cite{VR-PRB11} is that there is no particular reason that the glassy state obtained in the initial state after a particular thermal history be the same as the one obtained after a mechanically activated plastic event. In particular this choice is expected to allow rejuvenation and impact shear-banding~\cite{Debenedetti-PRL00,Varnik-PRL03,Varnik-JCP04}. In the following two variants of the model will be used depending on which of the rejuvenation or the disordered scenario is chosen. 

{\bf Elementary event} \-- Everytime the plastic criterion is
satisfied in a cell $(i,j)$, the basic brick of the model thus performs the following sequence of three operations:
\begin{itemize}
\item The strain field is incremented by a local plastic slip
$\delta\varepsilon_p$ drawn from a random distribution
$q(\delta\varepsilon_p)$;
\item The stress field is incremented by the internal stress field
  $\delta\sigma=G*\delta\varepsilon_p$ induced in the material by the local
  slip $\delta\varepsilon_p$. 
\item Finally the local plastic threshold $\sigma_c^{ij}$ is replaced by a new random value drawn in a {\it renewal}  distribution $p_r(\sigma_c)$.
\end{itemize}

{\bf Synchronous quasi-static dynamics} \--- The elastoplastic behavior of an amorphous material consists of series of elastic loading branches interspersed of stress drops. Starting from a stable configuration, the weakest site $(i^*,j^*)$ is identified as the one that minimizes the plastic strength field $x_{ij} = \sigma^c_{ij} - \sigma_{ij}$. A plastic event is performed at $(i^*,j^*)$. The elastic stress redistribution has then potentially destabilized a set of other cells. After identification, this set of unstable cells is updated synchronously~\cite{Tyukodi-PRL18}. Again, the series of plastic events may have destabilized other cells. This operation of identification of unstable sites and synchronous updates is performed until the avalanche stops and a stable configuration is recovered. The whole sequence is then repeated until a target macroscopic strain value is reached.

This series of operations thus defines a quasi-static driving in the very same spirit as
AQS in atomistic simulations: a global homogeneous deformation is
applied up to the first instability takes place. The first plastic
event triggers series of successive plastic events (an avalanche). The
global level of deformation is thus kept constant up to the end of the
avalanche. These two steps (loading up to the end of the elastic
branch, plastic avalanche) are repeated as long as
desired~\cite{Tyukodi-PRL18}.

\subsection{Two Dimensionless parameters }


Before addressing in the next section the implementation of our coarse-graining strategy we briefly discuss here the role of two dimensionless parameters, associated to disorder strength and rejuvenation, respectively.  The stationary regime is expected to be controlled solely by the disorder strength while the description of the transient regime ({\it e.g.} stress peak) requires in addition the rejuvenation parameter.

{\bf Disorder strength} \--- In analogy with depinning we can define a ``disorder strength'' $s$ that balances the  fluctuations of the local yield stress $\Delta \sigma^c$ with the stress drop $\Delta \sigma^{in}$ induced by a  plastic event of amplitude $\varepsilon_0$:  

\begin{equation}
        s = \frac{\Delta \sigma_c^r}{\Delta \sigma^{in}}\\
      =  \frac{\Delta \sigma_c^r}{\mu^* \varepsilon_0}\;,
\end{equation}
where $\mu^*=\mu/(1+\mu/K)$ is an effective shear modulus associated with the stress drop within a plastic inclusion. 

In the depinning context, the strength disorder would control the transition between weak and strong pinning (see Ref.~\cite{PVR-PRL13} for a discussion in the case of crack front propagating in a heterogeneous material). In particular it controls the value of the effective toughness of the material (the depinning threshold).

In the present case of amorphous plasticity, the effect of this strength disorder is more subtle and is at least two-fold. If one considers plastic deformation confined within a thin  shear-band along the shear direction, one recovers the standard case of a depinning line with a positive restoration stress along the band. However if one considers a  deformation in the full system we get a stress interaction whose sign fluctuates depending on the polar angle so that far from the inclusion it can be regarded as a mechanical noise. In this context, as early discussed in Ref.~\cite{TPVR-Meso12} the disorder strength both controls the value of the flow stress $\sigma_F$ :
\begin{equation}
    \frac{\sigma_F - \overline{\sigma_c^r}}{\overline{\sigma_c^r}}
    = f(s) \;,
\end{equation}
and the diffusion properties upon deformation {\it via} the typical value of the strain associated to a plastic event $\varepsilon_0$
\begin{equation}
    \delta \varepsilon^2_{pl} = g(s) \varepsilon_{tot} \;.
\end{equation}
where $\varepsilon_{tot}= \varepsilon_{el} + \varepsilon_p$ is thetotal strain.

Note that both the disorder strength parameter $s$ and the reduced flow stress are defined with respect to the mean plastic threshold of the renewal distribution. Again, this parameter is intended here to control the stationary flow regime.

{\bf Rejuvenation/age parameter} \--- As mentioned above, when dealing with the disorder landscape of the amorphous solids,we have to consider i) the distribution $p_i(\sigma_c)$ of local yield stresses $\sigma_c^i$ in the initial configuration, which depends on the particular preparation protocol used to prepare the glass; ii) the  renewal distribution $p_r(\sigma_c)$  of the new local yield stresses $\sigma_c^r$ of the zones that rearranged. This allows to define an age (or rejuvenation) parameter:

\begin{equation}
    r = \frac{\overline{\sigma_c^i} - \overline{\sigma_c^r}}{\overline{\sigma_c^r}}
\end{equation}

A positive value of $r$ corresponds to an {\it aged} material that {\it rejuvenates} upon shear loading. In Refs.~\cite{VR-PRB11} we could show that such positives values of the age parameter were associated to shear banding: the larger $r$, the thinner and the more persistent the shear band. 

In the following, rather than performing a full parametric study, we restrict ourselves to two limit cases. In the first one, we use for the renewal distribution of local yield stress the very same one as in the initial state. In the absence of rejuvenation we thus keep $r=0$ and the only expected evolution of the actual distribution of local plastic thresholds is the statistical hardening induced by the exhaustion of weak spots upon loading~\cite{BVR-PRL02,TPVR-Meso12,Patinet-PRE20-rejuvenation}. In the second case, we test the rejuvenation scenario proposed in Ref.~\cite{Patinet-PRE20-rejuvenation}, {\it i.e.} we use for the renewal distribution, the distribution of thresholds associated to an instant quench from a supercooled  liquid at a temperature close to the mode coupling transition.  We thus use the distribution of local yield stresses obtained for the glass coined as MCT in section~\ref{section:AQS}. In this scenario, performing the numerical application from the data obtained for the distributions of local yield stress for the different glasses, we get for the age parameter:
\begin{align}
    r^{ESL} &= \frac{\overline{\sigma_c^{ESL}} - \overline{\sigma_c^{MCT}}}{\overline{\sigma_c^{MCT}}} \approx 0.14 \\
     r^{GQ} &= \frac{\overline{\sigma_c^{ESL}} - \overline{\sigma_c^{MCT}}}{\overline{\sigma_c^{MCT}}} \approx 0.65 
\end{align}

\section{Upscaling strategy\label{section:upscaling}}

We discuss here how to connect the above defined mesoscopic model with
the atomistic simulations. We identify the parameters of the models
from the results obtained at atomic scale but we also try to make more
explicit the (numerous) hypotheses associated this upscaling operation
that involves a drastic reduction of degrees of freedom. We finally
list the observables that will be compared between the simulations
operating at atomic and mesoscopic scale repectively.

Before going further, let us first emphasize an obvious statement: the present coarse-graining operation is intended to fail, the model being too simple and caricatural to expect a complete quantitative agreement. Still, it
is of interest to study to what extent this class of scalar models is
able to cope for the phenomenology of amorphous plasticity. In the
following we discuss the various coarse-graining parameters that we
can extract from atomistic simulations.

\subsection{Coarse-graining parameters}

\begin{itemize} 
\item Elastic properties \--- We use the mean shear and bulk moduli $\mu$ and $K$  measured in the initial state.  In  particular the spatial fluctuations of the elastic properties as well as their evolution upon deformation are disregarded.

\item Coarse-graining length scale \--- In the atomistic simulations  the plastic properties are estimated in disks of radius $\xi=5\sigma$. The systems under study have a linear size $L\approx 100  \sigma$. Here we use for the mesh size of the lattice model  $\ell=\xi\sqrt{\pi}\approx 8.86$, slightly larger than $\xi$ such that the  area of one cell coincides with the area of the disk ${\cal A}= \pi \xi^2$ used to estimate the plastic thresholds in atomistic simulations~\cite{PVF-PRL16,Patinet-PRE18}. The linear size of the box used to perform the atomistic simulations is thus about $11 \ell$. In the following most of the mesoscopic simulations are performed on small systems of size N=16.
  
\item Distributions of plastic thresholds \--- The distributions  $p_i(\tau^c)$ used to model the two glasses at mesoscopic scale in  their as-quenched initial state are directly the ones computed from  atomistic simulations using Refs.~\cite{PVF-PRL16,Patinet-PRE18}:
  \begin{equation}
    p^{G}_i(\sigma_c) = {\cal P}^{G}(\sigma_c)\;, 
  \end{equation}
where ${\cal P}$ stands for the distribution measured from atomistic simulations and the superscript $G=ESL,GQ$ refers to the glass considered.
\begin{figure*}[!ht]
 \begin{center} 
\begin{subfigure}[b]{0.9\columnwidth}
\includegraphics[width=\columnwidth]{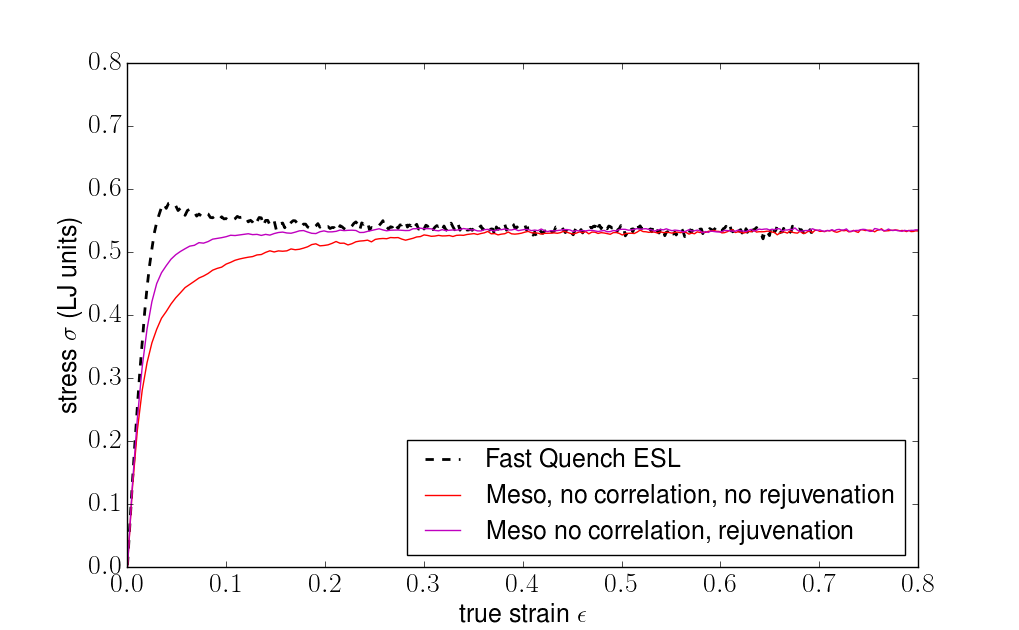}\caption{}
\end{subfigure}
\begin{subfigure}[b]{0.9\columnwidth}
\includegraphics[width=\columnwidth]{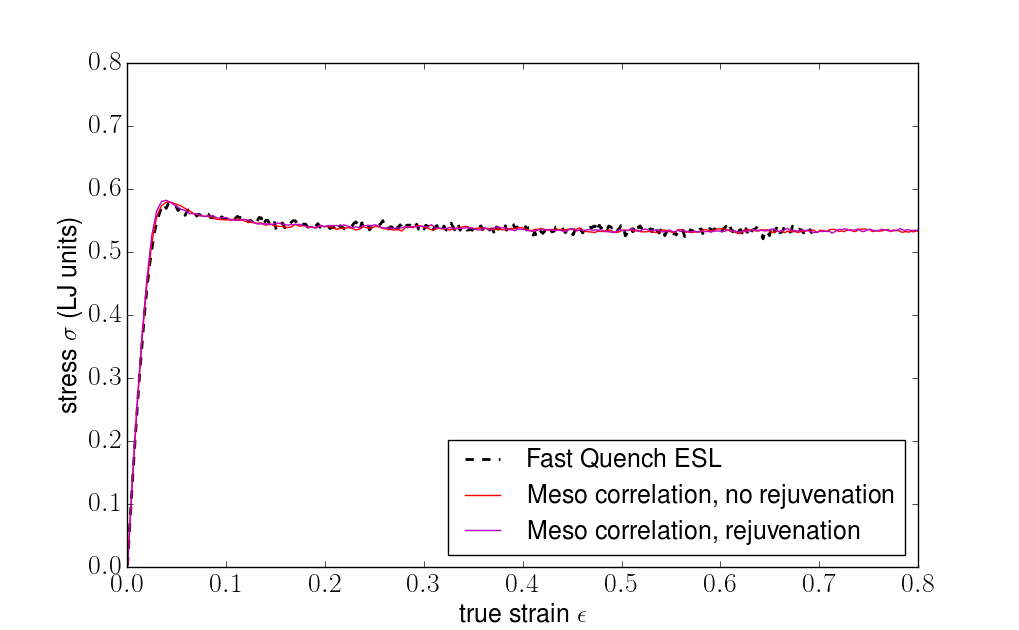}\caption{}
\end{subfigure}
\end{center}
  \caption{Stress-strain curves for the glass ESL and a mesoscopic model obtained without (a) and with (b) assuming correlation between the characteristic slip increment and the stress threshold.}
  \label{fig:AQS-vs-Meso-T0-stress-strain-curve}
\end{figure*}

\begin{figure*}[!ht]
\begin{center} 
\begin{subfigure}[b]{0.9\columnwidth}
\includegraphics[width=\columnwidth]{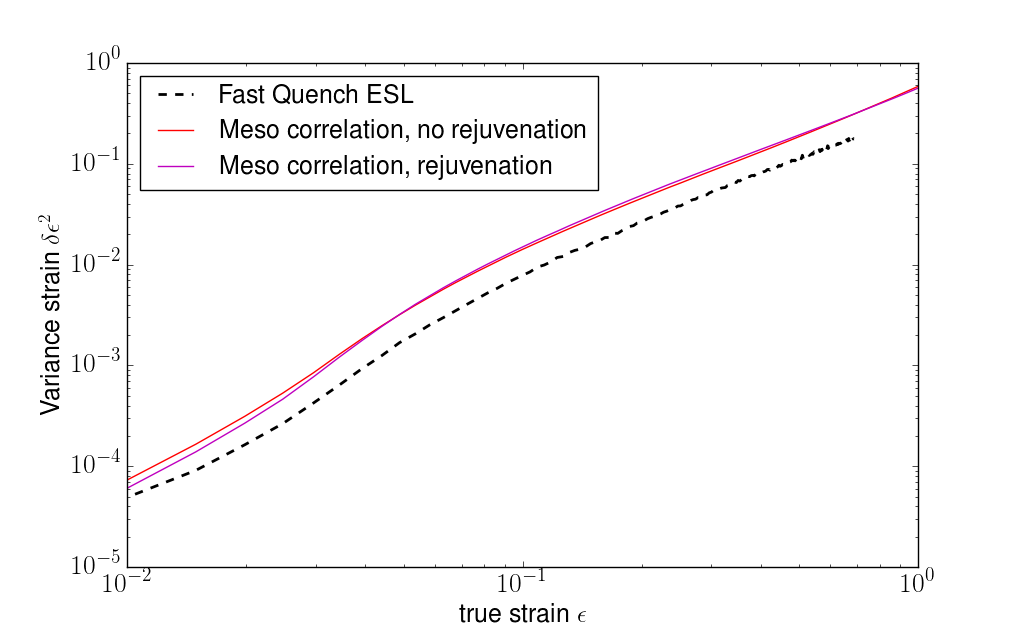}\caption{}
\end{subfigure}
\begin{subfigure}[b]{0.9\columnwidth}
\includegraphics[width=\columnwidth]{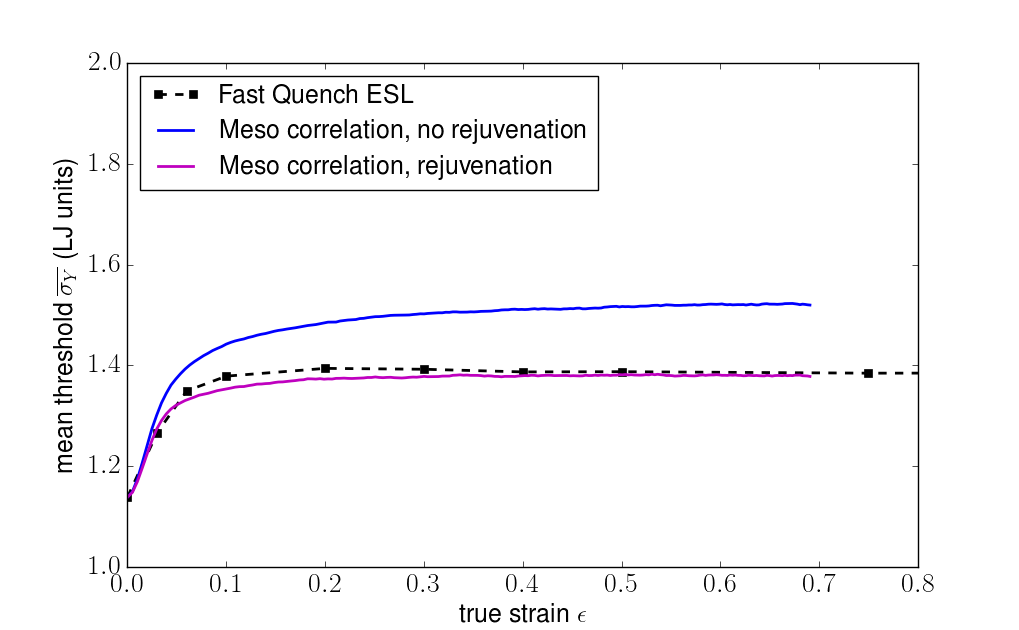}\caption{}
\end{subfigure}
\end{center}
  \caption{Variance of strain and stress vs true strain (a) in the ESL glass and in the meso model (b). Evolution of the mean threshold vs true strain in the ESL glass and in the meso model with and without rejuvenation.}
  \label{fig:fluctuations-ESL-vs-Meso}
\end{figure*}
In the first {\em disordered} variant of the model, we assume that  the distribution of plastic thresholds remains invariant under   deformation. We thus use $p_r(\sigma_c)=p_i(\sigma_c)$ for the  distributions of renewed plastic thresholds that characterize the  glassy state obtained after a plastic rearrangement. 
In the {\em rejuvenation} variant of the model, we account for a    deformation-induced rejuvenation effect. More precisely, we assume that the glassy local structure  obtained after a plastic rearrangement is specific, and {\it a priori} different from that of the as-quenched glass. We actually expect  that this new glassy structure contributes to the gradual formation   of the stationary glass eventually obtained after large  deformation. In that perspective, the distribution $p_r(\tau^c)$ of  the new plastic thresholds should be independent on the initial   condition and in the general case rather depend on the conditions   of the loading (strain rate, temperature). Following the  observation reported above we assume in this glassy variant that   the renewal distribution of thresholds is the same as the    distribution of initial thresholds in a glass abruptly quenched from a temperature close to the mode coupling transition:
     \begin{equation}
    p^{G}_r(\sigma_c) = {\cal P}^{MCT}(\sigma_c)\;.
     \end{equation}

\item Distributions of local slip increments \--- The amplitude of the   slip increments directly controls the level of elastic interactions   in the system and as such, the final value of the flow stress~\cite{TPVR-Meso12} or the broadening rate of   shear-bands~\cite{VR-PRB11}. This coupling  parameter~\cite{Budrikis-NatComm17} is thus crucial for the plastic   properties of amorphous solids. The determination of the local slip events remains extremely challenging~\cite{Rodney-PRL09}. The  successful mapping recently obtained by Albaret et  al.~\cite{Tanguy-PRE16} between the stress field measured in  atomistic simulation of amorphous silicon under shear and a sum of  elementary Eshelby stress fields opens the way toward a quantitative  analysis of these local events. In section~\ref{AQS-results} where we summarized results presented in Ref.~\cite{Patinet-PRE18}, the  local slip amplitudes were estimated after the measurement of the local stress drops obtained after the relaxation of the local glassy   structure following a plastic rearrangement. Local slip   amplitudes were shown to follow a power law distribution bounded by an exponential cut-off and to
  be significantly correlated with the local yield stress. 

In most mesoscopic models of amorphous plasticity, slip increments
  are settled at a constant value or drawn from a random distribution,
  generally uncorrelated with the plastic thresholds. In contrast,
  models based on an energy landscape naturally introduce a strong
  correlation between local slips and thresholds. The direct
  relationship between the two quantities thus depends on the precise
  shape of the potential energy landscape chosen in the
  model~\cite{Jagla-PRE17,Jagla-PRE18}.

  In the present model we consider for the slip increments an
  exponential distribution where we account for a direct dependence
  on the stress threshold:
\begin{equation}
    p(\delta\varepsilon_p | \sigma^c) = \frac{1}{\varepsilon_0(\sigma_c)}\exp \left(-\frac{\delta\varepsilon_p}{\varepsilon_0(\sigma_c)} \right) \;,
    \label{eq-distrib-slip}
    \end{equation}
    
where the mean value of the slip increment obeys a parabolic
  dependence to the local yield stress:
  \begin{equation}
    \quad \varepsilon_0(\sigma_c) = c_0 +c_1 \sigma_c^2\;.
    \label{eq-slip-vs-threshold}
     \end{equation}
\end{itemize}

In the following, the coefficients $c_0$ and $c_1$ will be used as
tuning parameters to reproduce as much as possible the stress-strain
curves (the values of the maximum stress and of the flow stress). In order to test the importance of the correlation between local stress drops and local yield stresses, we will also try to fit only the flow stress value through the tuning of the sole coefficient $c_0$, thus fixing the other coefficient $c_1=0$.  

As detailed below, a systematic comparison will then be made between atomistic and mesoscopic results for different observables of interest: strain field, stress field and their fluctuations. Note again that most of the parameter being directly extracted from the initial configurations of the atomistic simulations, we are left with a very constrained problem so that we deal only with one or two fitting parameters to reproduce the results.

\subsection{Observables}

The extreme simplicity of the mesocopic model obviously limits the extent of observables to be compared between atomistic and mesoscopic scales. In the following, we present the evolution of the stress and strain fields upon shearing as well as the evolution of the local plastic properties.

In order to make the comparison as quantitative as possible, we compute averaged behaviors (mean values or mean profiles) as well as their fluctuating part (variance and higher moments). In addition to the stress and stress fields we also give a special attention to the field of local yield stresses.

\section{Deformation of a soft/fresh glass}\label{section:ESL}

We consider here the deformation of the soft glass ESL. As discussed in the previous section, we want to gradually enrich the description of the disordered landscape and test the importance of two hypotheses: a rejuvenation effect and the correlation between local thresholds and slip increments.

\subsection{Stress-strain curve}

In the most naïve version, the EPM considers no correlation and no rejuvenation. We implement a distribution of plastic thresholds similar as the one measured by the local yield stress method for the glass ESL. When the plastic criterion is satisfied in a cell, the latter experiences a local slip. The amplitude of the slip is taken from an exponential distribution of width $c_0$. To account for a change of local structure a new plastic threshold is drawn from the very same distribution as the initial one. Here $c_0$ is a tunable parameter. Its value $c_0=0.0626$ is adjusted so that the stress plateau of the stress-strain curves reproduces the results of the AQS atomistic simulation.

In a second version, rejuvenation is considered: when a site has slipped, the plastic threshold is renewed but its value is now taken from a rejuvenated distribution. Here, following the scenario discussed in Ref.~\cite{Patinet-PRE20-rejuvenation}, we use the distribution of thresholds obtained for a glass instantly quenched (inherent structure) from a liquid equilibrated at $T=0.38$, a temperature close to the Mode-Coupling transition. As in the previous case, the width $c_0$ of the distribution of slip increment is used as a tunable parameter. With $c_0=0.0506$ to reproduce the level of the stress plateau of the stress strain curve.

We show in Fig.~\ref{fig:AQS-vs-Meso-T0-stress-strain-curve} the comparisons of the stress-strain curve obtained with atomistic and mesoscopic simulations, respectively. In both cases, the stationary regime is nicely reproduced. In other words, the stress plateau can be reproduced with various distributions of local yield stresses. We see however that in the absence of correlation  between local thresholds and slip increments (Fig.~\ref{fig:AQS-vs-Meso-T0-stress-strain-curve}a), the mesoscopic models fail to reproduce the transient part of the stress-strain curve. In particular, while a discrete peak is observed in atomistic simulations, the stress-strain curve remains monotonic in the mesoscopic case. The account of rejuvenation slightly improves the results but the one-parameter fit performed here to reproduce the stress plateau clearly does not allow us to get the transient part of the plastic behavior. 

In the presence of correlation between local thresholds and slip increments (Fig.~\ref{fig:AQS-vs-Meso-T0-stress-strain-curve}b), we observe on the contrary that the mesoscopic simulations reproduce very nicely the atomistic stress strain curve. Here two tunable parameters were used. The width $\varepsilon_0$ of the exponential distribution used to generate a slip increment $\delta\varepsilon_p$  after a threshold $\sigma_c$ is given by $\varepsilon_p=c_0+c_1\sigma_c^2$. The additional parameter $c_1$ thus quantifies the level of correlation. The pair of parameters $(c_0,c_1)$ is tuned so that the mesoscopic stress-strain curve reproduces the level of the stress peak and that of the stress plateau. We observe that the full curve is nicely fitted. This close agreement persists independently on the choice to use  or not the hypothesis of rejuvenation. We get $(c_0,c1)=(0.0304,0.0216)$ in the absence of rejuvenation and $(c_0,c1)=(0.0276,0.0206)$ with rejuvenation. These first results thus tend to emphasize the importance of the underlying structure of the potential energy landscape (PEL). In order to reproduce the atomistic results we need to consider a hierarchical-like structure of the PEL such that the deeper the well (the larger the threshold), the further the neighboring well (the larger the  slip). 


Interestingly, the additional coefficient $c_1$ does not significantly affect the flow stress. If one computes the averages value $\overline{\varepsilon_0} = c_0 + c_1 \overline{\sigma_c^2}$, we get  $\overline{\varepsilon_0} = 0.0634$ and $0.0521$ in the presence of correlations, {\it i.e.} very close values of the ones obtained without correlation. In spite of the additional complexity the dimensionless parameter $s$, directly proportional to $\varepsilon_0$ thus fully controls the level of the plateau stress. 

\subsection{Strain fluctuations}


In Fig.~\ref{fig:fluctuations-ESL-vs-Meso}a we show the evolution of the strain variance upon loading. Here we considered correlation between stress drops and local yield stress and used the coefficients $(c_0,c_1)$ fitted to reproduce the stress-strain with and without rejuvenation. 
We observe that while the fluctuations obtained in the mesoscopic simulations are overestimated by a factor about $1.5$  with respect to the atomistic results, the global evolution of the variance upon loading is very nicely reproduced in the early regime as well as in the late regime. We remark that the level of fluctuations apparently does not depend on the rejuvenation hypothesis: the curve almost superimpose. Again, as proposed in Ref.~\cite{TPVR-Meso12} this indicates that the diffusive behavior of the strain field is fully controlled by the sole disorder strength parameter {\it via} the typical value of the plastic slip $\varepsilon_0$



\subsection{Local yield stress}
In Fig.~\ref{fig:fluctuations-ESL-vs-Meso}b we show the evolution of the local yield stresses upon loading. The atomistic results show a clear hardening trend. The average local yield stress shows an increase of about 10\% upon loading. This increase takes place in the transient regime of deformation so that a stationary value is quickly reached. As for the strain fluctuation, we consider here the correlation case and study the effect of rejuvenation. Without rejuvenation, the mesoscopic model qualitatively reproduces the hardening trend, but tends to to overestimate the plastic thresholds. However with rejuvenation, one observe a quasi-quantitative agreement with the atomistic results. In contrast to the plateau stress and the strain fluctuation, the reproduction of this observable, thus crucially depends on the rejuvenation hypothesis. Accounting for the fact that the level of rejuvenation is not a free parameter but has been arbitrarily fixed by considering the renewal distribution as deriving from the glass MCT, this results gives a strong argument to the validity of the rejuvenation scenario.


\section{Shear-banding of a hard/gradually quenched glass}\label{section:GQ}

We consider here the deformation of the hard gradually quenched glass GQ. As discussed in the previous section, we want to gradually enrich the description of the disordered landscape. We first focus on the respective effects of correlation and rejuvenation on the stress-strain curves. Since the latter exhibits a stress peak in association with a shear-banding behavior, we then follow the evolution of a localization ratio based on the participation ratio of the strain field. As for the fresh glasss ESL we also follow the evolution of the mean value of the local yield stress field.



\subsection{Stress-strain curve}

\begin{figure*}
\begin{center}  
\begin{subfigure}[b]{0.9\columnwidth}
\includegraphics[width=\columnwidth]{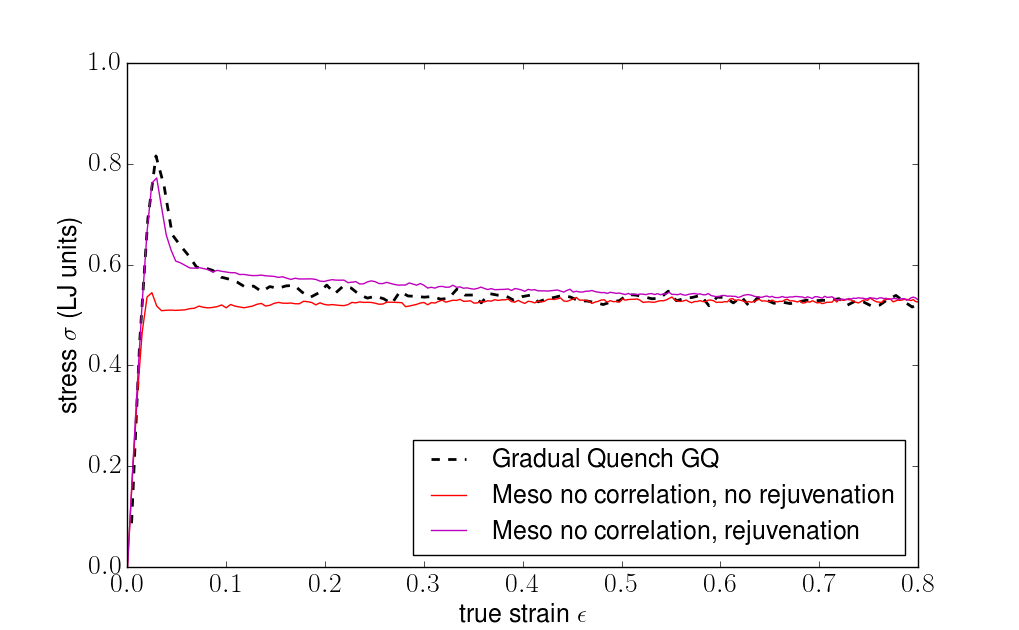}\caption{}
\end{subfigure}
\begin{subfigure}[b]{0.9\columnwidth}
\includegraphics[width=\columnwidth]{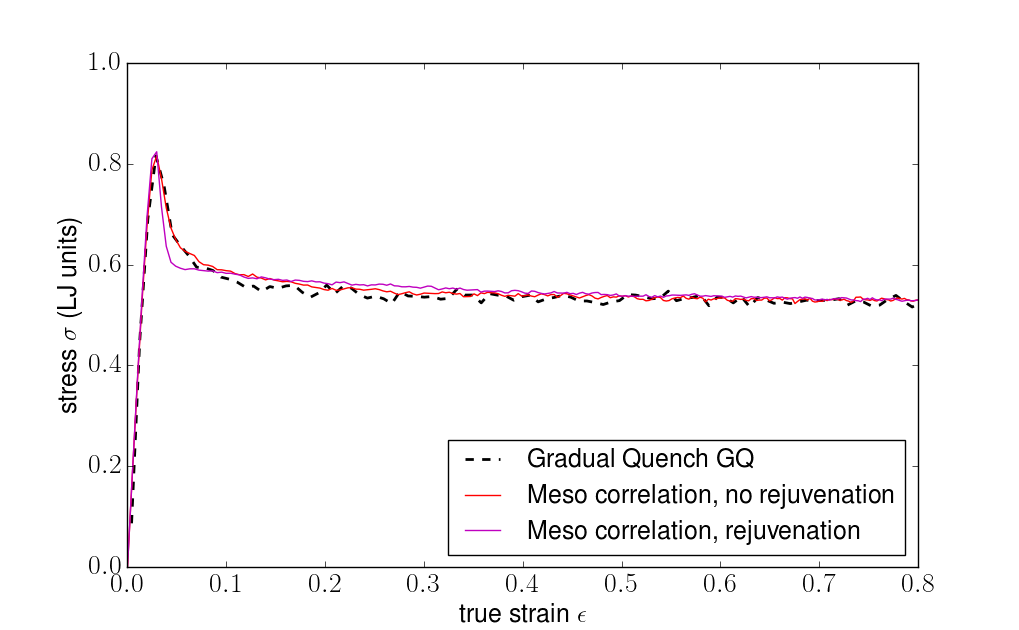}\caption{}
\end{subfigure}
\end{center}
\caption{Stress-strain curve  in the GQ and in the meso model with and without correlation between local stress drops and yield stress}
 \label{fig:stress-strain-GQ-vs-Meso}
\end{figure*}

We show in Fig.~\ref{fig:stress-strain-GQ-vs-Meso} the comparisons of the stress-strain curve obtained for the hard glass GQ with atomistic and mesoscopic simulations, respectively. As for the soft glass ESL, Fig.~\ref{fig:stress-strain-GQ-vs-Meso}a shows the results in the absence of correlation between stress drop and local yield stress. In the absence of correlation and rejuvenation {($c_0=0.0836$)}, the plateau stress is correctly reproduced but the only trace of a stress peak is a tiny bump. Conversely, in the presence of rejuvenation  {($c_0=0.02$) }, despite the fact that the fitting procedure operates only on the stress plateau value, a well defined stress peak gets clearly visible and almost reaches the atomistic value. {The general shape of the stress-strain curve shows however clear discrepancies compaed to atomistic data}.

When accounting for the correlation between stress drops and local yield stresses and performing a two-coefficients fitting procedure, the results get better. As required, both the value of the plateau stress and the stress peak are nicely reproduced. In the absence of rejuvenation we get $(c_0,c1) = (0.0344,0.0164)$ and a very good agreement in the stress strain curve. In the presence of rejuvenation we get  $(c_0,c1) = (0.0235,0.0163)$. Interestingly, the global reproduction of the stress-strain curve is not so good in this case with stress peak significantly sharper than in the atomistic case. As for the fresh glass ESL we note that the mean value of the slip increment $\overline{\varepsilon_0}$ keeps almost constant with and without correlation. Without rejuvenation we get  $\overline{\varepsilon_0} = c_0 + c_1 \overline{\sigma_c^2} = 0.0833$ and with rejuvenation  $\overline{\varepsilon_0}= 0.0429$. Again these values are extremely close to those obtained in the absence of correlation, which confirms that the level of the stress plateau is fully controlled by the value $\varepsilon_0$ and/or dimensionless parameter $s$ (strength of disorder).


\subsection{Localization}

In Fig.~\ref{fig:localization-seuils-GQ-vs-Meso}a we show the evolution of a localization index $LOC$ of the strain field upon shear loading. This index is based on the participation ratio $PR$ classically used to characterize localization phenomena. In practice, for a strain field $\varepsilon_{ij}$ on a grid of linear size $N$, we define
\begin{equation}
    LOC(\varepsilon) = \frac{1}{N. PR(\varepsilon)}= 
    \frac{\sum_{ij} \varepsilon_{ij}^4}{(\sum_{ij} \varepsilon_{ij}^2)^2}
\end{equation}
In the case of a localization of a single band, we get $LOC(\varepsilon) = 1$ while for a fully homogenous field we get  $LOC(\varepsilon) = 1/N$. The glass GQ shows a strong localization behavior. The maximum value is obtained just after the location stress peak and corresponds to the nucleation of a shear band. The subsequent broadening of the band then induces a gradual decrease of the localization index. 

We compare these results with mesocopic simulations performed with correlation of stress drops with local yield stress and we discuss the effect of the rejuvenation hypothesis. The localization  behavior appears to be very well reproduced by the mesoscopic model with rejuvenation and correlation. The global trend is recovered: a fast increase associated to the nucleation of the band, followed by a slower decrease associated to the broadening of the band. The maximum value of the localization index is reasonably reproduced.

Note that even in the absence of rejuvenation, we still get a significant localization behavior. The global trend is recovered but the level of localization remains under-estimated.

\subsection{Local yield stress}

\begin{figure*}
\begin{center}  
\begin{subfigure}[b]{0.9\columnwidth}
\includegraphics[width=\columnwidth]{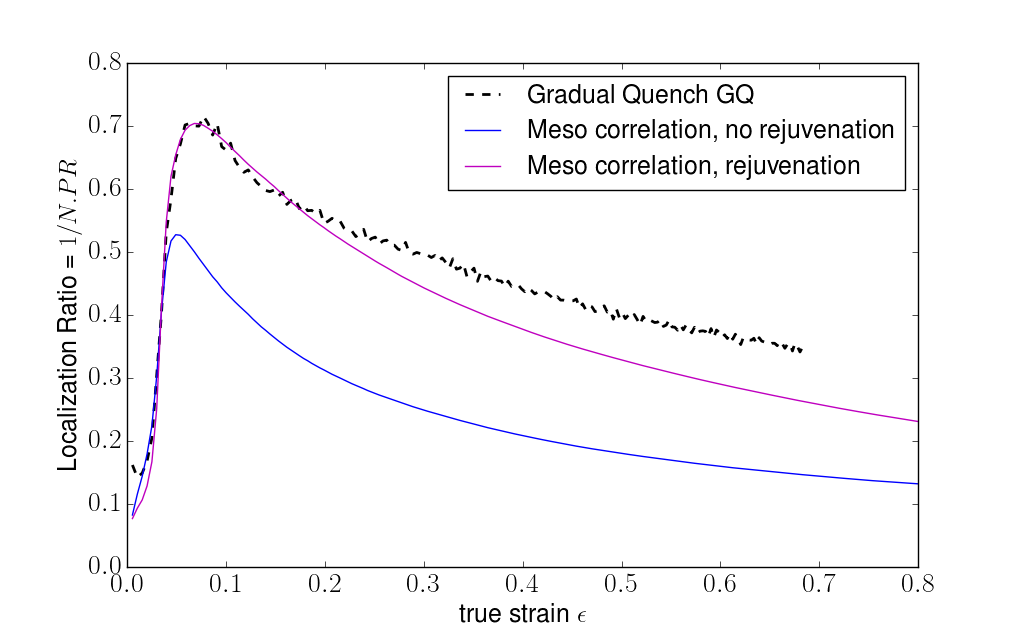}\caption{}
\end{subfigure}
\begin{subfigure}[b]{0.9\columnwidth}
\includegraphics[width=\columnwidth]{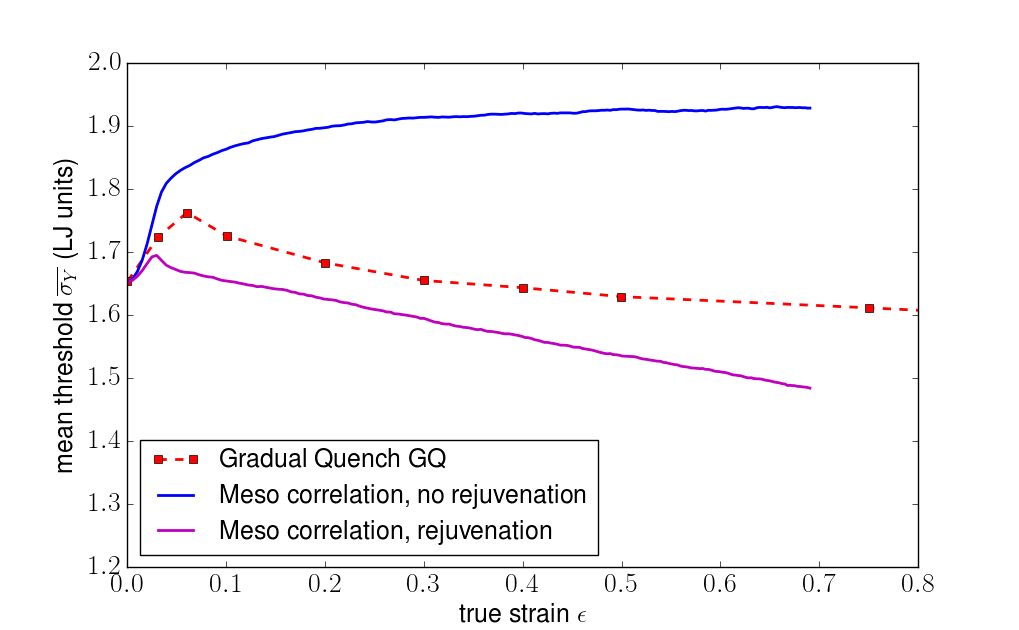}\caption{}
\end{subfigure}
\end{center}
  \caption{Evolution of the localization ratio and of  the mean threshold vs true strain in the GQ glass and in the meso model with and without rejuvenation.}
  \label{fig:localization-seuils-GQ-vs-Meso}
\end{figure*}

In Fig.~\ref{fig:localization-seuils-GQ-vs-Meso}b we show the evolution of the mean value of the local yield stress upon shear deformation. As discussed in Ref.~\cite{Patinet-PRE20-rejuvenation}, a rather complex behavior emerges. we observe first a slight increase of the mean plastic threshold that can be associated to a statistical hardening process. Then a shear-band nucleates and broadens. The associated replacement of the old hard glass by a fresh softer one induces a gradual decrease of the average local yield stress. 

We compare these atomistic results with the mesoscopic ones obtained with and without rejuvenation. Without rejuvenation, there is actually no mechanism at play that could tend to decrease the mean local yield stress. We thus only observe a growing trend which results from the statistical hardening discussed in section~\ref{section:ESL} in the case of the glass ESL and the mesoscopic model significantly overestimates the values of the atomistic mean yield stress. 

When accounting for rejuvenation, the mesoscopic results get better. In particular, we recover a similar trend as with the atomistic simulations: a first increasing regime, followed by a slow decreasing regime. However form the quantitative aspect, the agrrement is not spectacular and the mesoscopic model tends here to slightly underestimate the atomistic results. Still, it clearly appears here that the rejuvenation hypothesis is crucial to recover at least qualitatively the complex evolution of the mean yield stress upon loading.

\begin{figure*}[!ht]
\begin{center}  
\begin{subfigure}[b]{\columnwidth}
\includegraphics[width=0.9\columnwidth]{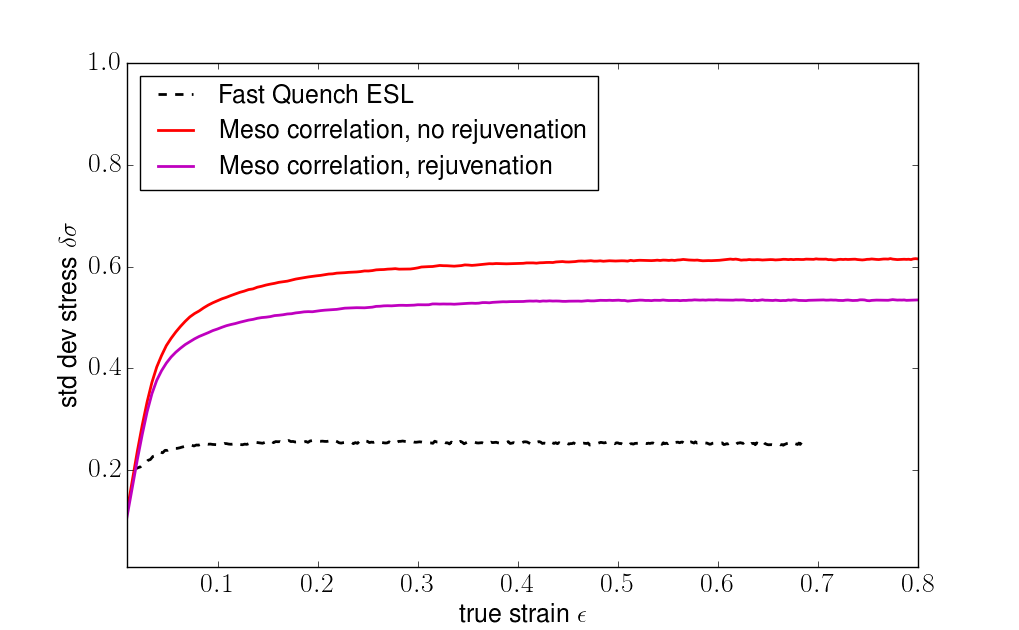}\caption{}
\end{subfigure}
\begin{subfigure}[b]{\columnwidth}
\includegraphics[width=0.9\columnwidth]{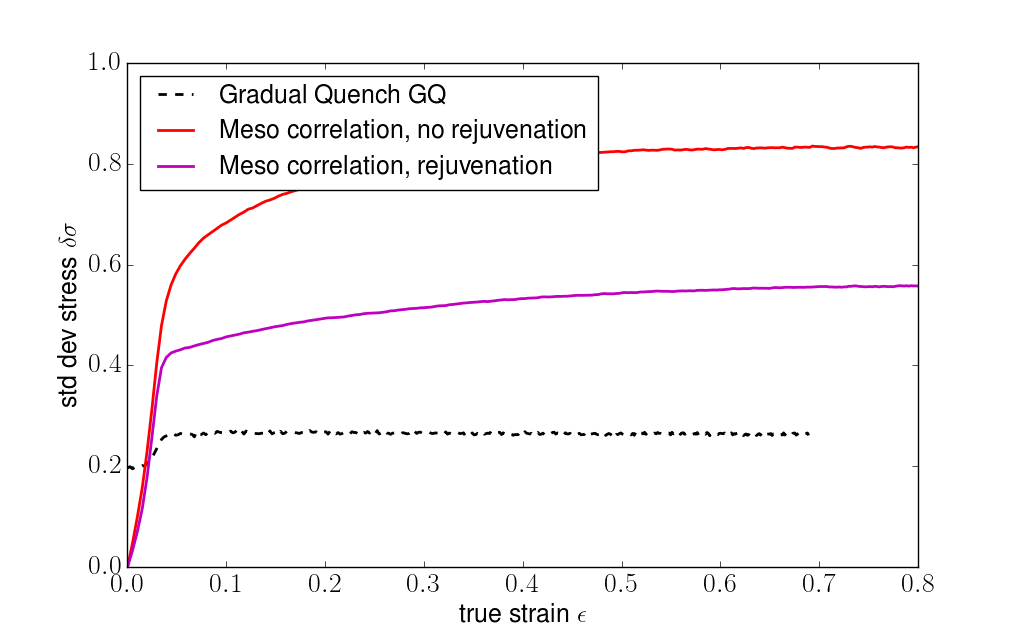}\caption{}
\end{subfigure}
\end{center}
  \caption{Evolution of the standard deviation of the internal stress field in the glasses ESL and GQ vs meso model.}
  \label{fig:internal-stress-fluctuations-ESL-GQ-vs-Meso}
\end{figure*}

\section{Discussion}\label{section:Discussion}

In this contribution, we discussed the coarse-graining of amorphous plasticity from the atomistic to the mesoscopic scale. We deliberately restricted ourselves to a very simplistic scalar version of a mesoscopic elasto-plastic model. We tried to identify a few key parameters of the underlying glassy landscape from the failures and successes encountered in this approach. In particular, we tried to test the rejuvenation scenario proposed in Ref.~\cite{Patinet-PRE20-rejuvenation}.

We emphasized the role of two dimensionless parameters, respectively associated with the strength of the disorder with respect to the typical stress drop and with the relative age of the initial glass with respect to the fresher one that emerges upon loading. 

Testing our approach on two glasses of contrasting soft and hard behaviors, we could nicely reproduce the stress-strain curves, the evolution of the mean plastic thresholds and the localization behavior. We also got a qualitative agreement for the diffusive behavior of the strain field of the glass ESL.

\subsection{Structure of the disordered yield stress landscape}

The results we obtain are consistent with the rejuvenation scenario proposed in Ref.~\cite{Patinet-PRE20-rejuvenation} according to which, upon shear deformation, the initial glass is gradually replaced by a fresh new one whose mechanical properties are close to a glass abruptly quenched from a temperature close to the mode coupling transition. We also made evidence for the necessity of correlating the local stress drops following a local plastic event with the height of the stress barrier, {\it i.e.} the local yield stress. Such a relation is not unexpected and directly derives from the structure of the underlying potential energy landscape. Still, it is interesting to show that accounting for such a correlation between local stress drop and yield stress is necessary to reach a quantitative agreement between atomistic and mesoscopic simulations. Additional features of the disordered yield stress landscape certainly deserve further study. In particular, we did not consider any memory. In the present implementation of the model, once a plastic event takes place, the former yield stress is immediately forgotten, independently of the amplitude of the local re-arrangement. This memory effect and its dependence on the initial preparation has been recently discussed in Ref.~\cite{Patinet-Acta22}.

\subsection{Elastic heterogeneity and internal stress fluctuations}

In the present version of the mesoscopic model, elastic properties are homogeneous and do not evolve upon deformation. This assumption is at odds with the actual behavior observed in atomistic simulations. The macroscopic moduli significantly change upon shearing. In the stationary state, obtained after very large deformation, one gets for the model glasses under study $\mu_{ss}=13.2$ and $K_{ss}=59$ to contrast with the initial values $\mu_{ESL}=14.7$, $K_{ESL}=52$ and $\mu_{GQ}=19$, $K_{GQ}=55$. These are significant changes, up to a few tens of per cent for the glass GQ.

The spatial heterogeneity of the elastic properties was also completely neglected in the present study. From the qualitative point of view, this omission may not be dramatic since the structure, and the symmetry of the associated internal stress fields are likely to be close to that of the Eshelby stresses induced by plastic inclusions. From the quantitative point of view, the impact may be more significant.

Along with the relative (and sometimes impressive) success of the coarse-graining attempts reported so far in the present manuscript, it is essential to insist on the apparent failure that can be identified. In Fig.~\ref{fig:fluctuations-ESL-vs-Meso}, we reported strain fluctuations that overestimated the atomistic results by a factor of about 1.5. This overestimating trend of the microscopic fluctuations is all the more impressive when we focus on the internal stress field. In Fig.~\ref{fig:internal-stress-fluctuations-ESL-GQ-vs-Meso}, we report the evolution of the standard deviation of the internal stress fields for the ESL and GQ glasses and the associated mesoscopic models. While the latter does not account for the stresses induced by the elastic disorder, they overestimate the fluctuations by 2 to 3. The reasons behind this large discrepancy are not entirely clear at this stage. The scalar character of the model may induce an effect of "constructive interference", artificially enhancing fluctuation that would be partly smeared out by an orientational disorder of the weak slip directions (see Refs~\cite{Patinet-CRP21,Patinet-Acta22} for discussion of the effect of full tensorial disorder). Another source of discrepancy may be the overestimation of the local yield stress by the frozen matrix method. In the same spirit, the mesoscopic model relies on a hypothesis of continuum elasticity in the rearranging regions and disregards non-affine elastic effects. The impact of these different points on internal stress fluctuations certainly deserves further study.


\section{Acknowledgements}
BT acknowledges support from the Romanian Ministry of Education and Research, CNCS-UEFISCDI, project no. PN-III-P1-1.1-PD-2019-0236, within PNCDI III and the Brandeis MRSEC NSF DMR-2011846. DV and SP acknowledge countless discussions with S. Roux on mesoscopic models of amorphous plasticity and their connection with atomistic simulations.

\section*{References}
\bibliographystyle{iopart-num}
\bibliography{biblio_CG}

\end{document}